# Attaining High-performing Software Teams with Agile and Lean Practices: An Empirical Case Study

**Vaibhavi Oza, Petri Kettunen, Pekka Abrahamsson\*, Jürgen Münch**
*University of Helsinki*
*Department of Computer Science*
*P.O. Box 68 (Gustaf Hällströmin katu 2b), FI-00014, Finland*
*\*Free University of Bozen-Bolzano*
vaibhavi.oza@gmail.com, petri.kettunen@cs.helsinki.fi, pekka.abrahamsson@unibz.it,
juergen.muench@cs.helsinki.fi

## ABSTRACT

*This paper presents an empirical study on how self-organized software teams could attain high performance using agile and lean practices. In particular, the paper qualitatively examines characteristics of high performance and self- organization in one project team. The case under study is a customer-driven student project, carried out to develop an alpha-version prototype. The paper also studies how certain agile software practices aid in initialising self-organization in the team. The main results indicate that self-organization as supported by certain Agile and Lean practices helps teams in achieving higher performance.*

## 1 INTRODUCTION

Companies have been changing their approach to software development [6]. Achieving better performance and higher quality with lighter (agile) processes have been some of the key drivers. Self-organization in teams is considered to be one of the key factors to develop this capability of adaptation with sustainable high performance [1].

We investigate this by empirically studying whether agile software practices help to induce self-organization and thereby develop and support high-performing teams. In other words, our line of inquiry forms two research questions (RQ) as shown in Figure 1.1: RQ1) how agile practices help in achieving self-organization in the team and RQ2) is there a connection between high performance and self-organization in the team?

Empirical data for our study come from a customer project conducted at the Software Factory[1] setup at the University of Helsinki [9].

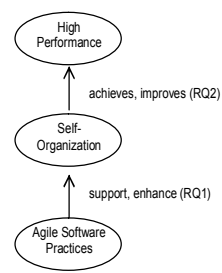

**Figure 1.1: Research hypotheses investigating connections between Agile/Lean software practices, self-organization and high performance**

We interviewed five team members who have worked in the project and also the customer of the project. In this paper, we focus on the team members perspectives on characteristics of self-organization, integration of agile practices in self-organization and how high performance is compatible with self-organization as stated in the research questions above.

## 2 LITERATURE REVIEW

One of the key factors for high performance has been recognized to be self-organization [10]. The literature also offers insights on how self-organization could be developed within the teams [8]. We integrate key works from the management literature and emerging studies from the software research in this section.

### 2.1. Characteristics of self-organized teams

Self-organization in the team appears quite straightforward from the outset but may become unmanageable, if not integrated carefully as demonstrated for instance by Ikonen [11].

Karhatsu [1] proposes key characteristics of self-organized teams in the context of agile software development. We take Karhatsu s framework (Fig. 2.1) as a baseline in this study for our empirical work. It includes key characteristics of self-organized teams.

---

[1] www.softwarefactory.cc

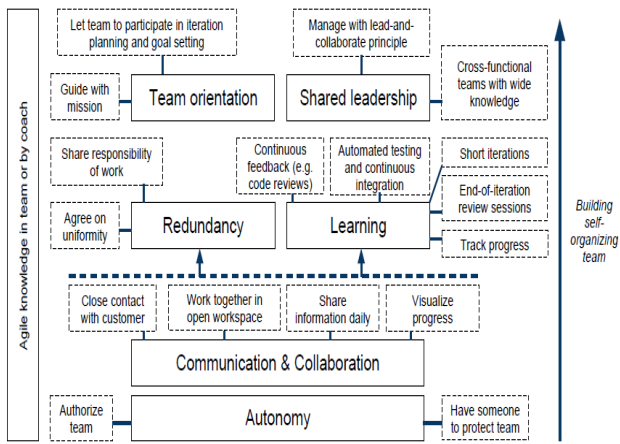

**Figure 2.1:** Theoretical framework depicting characteristics of self-organized teams with agile software development practices (after [1])

### 2.2. The team's high performance and self-organization

Team performance includes both individual results and what we call "collective work products"[2]. Performance can be identified by these "collective work products". We acknowledge some of the key characteristics of high performing teams from Hancock [3] and then later compare them with our study results.

Software organizations are always looking for ways to improve their team's performance. We assume five factors that may foster the improvement of a team's performance:

- Focus on the selection of core members [4]
- High congruence [5]
- To create an open space for knowledge sharing [4]
- Feedback [5]
- Establish a "goal-oriented incentive" mechanism [4]

High performance depends on the self-organizing capability of teams [7, 8]. As an example, self-organization can be achieved by assigning full authority to the teams [8]. It is noted that self-organization also spans across areas of responsibilities in management and work design dimensions.

### 2.3. Role of Agile software practices in self-organization and team performance

Agile practices such as Scrum and XP (Extreme Programming) have shown much evidence of how they help boost a team's performance in software development [7]. However, the detailed study or analysis of how agile practices enable self-organization seems to be missing. Our study builds on such existing works to better understand how agile practices can help develop self-organization and high-performing teams.

## 3 RESEARCH DESIGN

We conducted a qualitative empirical study about the research questions stated in Section 1 using open-ended thematic interviews. The qualitative patterns from the collective interview transcripts were then examined to conduct a thematic analysis.

We selected five of the ten Software Factory project *Planesweep*[2] team members for interviews. The selection of interviewees was based on participant observation of the first author.

We used semi-structured interviews to collect qualitative data. The interview questions and overarching objectives were the same for all the interviewees. It was, however, possible for interviewees to talk freely about a range of issues within the specific theme or question. The time taken for each interview varied from 50 to 90 minutes. In addition to the recordings, some written notes were taken during the interviews.

We used thematic analysis to analyze qualitative data. The analysis after the data collection consisted of systematic searches through each transcript, categorizing perspectives into self-organizing characteristics following the selected reference frame (Sect. 2.1).

## 4 RESULTS

In this section, we explain the qualitative integration of the results to reflect on our primary questions (see Section 1), i.e., 1) how agile practices help in achieving self-organization in a team and 2) whether high performance is related to self-organization in teams.

### 4.1. Agile/Lean practices aiding self-organization

The results indicate that the agile approach helped teams to self-organize. Agile/Lean practices also aided the team's ability to self-organize in different ways – including learning, autonomy, shared leadership, team orientation, redundancy, and communication-collaboration. As shown in Table 4-1, the importance of the Kanban board clearly emerged as one of the crucial practices in developing self-organizing characteristics in the team.

As another example, weekly retrospectives helped in achieving some of the self-organizing characteristics (Learning, Communication & collaboration). One interviewee mentioned:

> "Well Retrospectives were very useful and they weren't just complaining about problems but we also solved those problems. That was very good and daily meetings were useful but maybe Retrospectives were the best part of the process model I think."

---
[2] http://www.softwarefactory.cc/project/planesweep

**Table 4-1. Agile/Lean practices supporting case team self-organization**

| Self-organizing characteristics/ Method | XP | Scrum | Lean software development |
|---|---|---|---|
| Autonomy | | Backlog | |
| Team Orientation | Pair-programming | | Kanban Board |
| Shared leadership | Demo culture | | |
| Redundancy | Co-located teams | | |
| Learning | | Retrospectives | Kanban Board |
| Communication & collaboration | | Daily stand-ups, Sprint planning | Kanban Board |

## 4.2. High performance and self-organization

We examined how the characteristics of high-performing teams and self-organized teams were related in our qualitative data. Table 4-2 shows that most of the high-performing team characteristics have some connection to ⌐Autonomy⌐ and ⌐Shared leadership⌐. It was also noted that none of the high-performing team characteristics showed any connection to ⌐Redundancy⌐. We also see in Table 4-2 that there was no connection between ⌐A belief in shared aims and objectives⌐ and any of the self-organization characteristics.

**Table 4-2. Empirical evaluation of high performing vs. self-organization characteristics (X = supporting observations)**

| | Self-organized Team Characteristics [1] | | | | | |
|---|---|---|---|---|---|---|
| | A- Autonomy, T ⌐Team orientation, S- Shared leadership, R ⌐Redundancy, L ⌐Learning, C- Communication and collaboration | | | | | |
| High-performing team characteristics [3] | A | T | S | R | L | C |
| A belief in shared aims and objectives | | | | | | |
| A sense of commitment to the group | X | | X | | | |
| Acceptance of group values and norms | | X | | | X | |
| A feeling of mutual trust and dependency | X | | X | | | X |
| Full participation by all members and decision making by consensus | X | X | X | | | |
| A free flow of information and communications | X | | X | | X | X |
| Open expression of feelings and disagreements | X | X | X | | X | |
| The resolution of conflicts by the members themselves | X | | X | | | |

## 5 DISCUSSION AND CONCLUSION

We specifically found that Agile and Lean practices such as ⌐Retrospectives⌐, ⌐Daily stand-ups⌐ and Kanban had a strong influence on developing self-organization in the team (RQ1). These practices indicated improvement in ⌐Autonomy⌐, ⌐Shared leadership⌐ and ⌐Learning⌐. In relation to high-performing team characteristics (RQ2), ⌐Autonomy⌐ and ⌐Shared leadership⌐ turned out to have clear connections to them.

This is a preliminary empirical study. One of the biggest limitations is that the study has not measured the performance metrics of the team with respect to their correlation with self-organization or Agile practices.

Another major limitation we recognize is in the scope of the single case findings. Further work is planned to address larger projects also in distributed set-ups, which may be harder to self-organize.

In this study, we empirically examined how Agile/Lean software practices help in achieving self-organization in the team and if there is a connection between high performance and self-organization in the team.

We found that certain agile software practices had influence on developing self-organization in the team. With respect to high-performing team characteristics, autonomy and shared leadership turned out to have clear connections to them. However, our study requires more supporting cases to conduct empirical studies on similar inquiries in different contexts.

ACKNOWLEDGEMENTS

This work was supported by TEKES as a part of the Cloud Software program of Tivit (Finnish Strategic Centre for Science, Technology and Innovation in the field of ICT) and the SCABO project (no. 40498/10). The authors would also like to thank Marko Ikonen (University of Helsinki) for his comments, and the members of the Software Factory project team.